# Electric-field-induced superconductivity in electrochemically-etched ultrathin FeSe films on SrTiO$_3$ and MgO

J. Shiogai[1,#], Y. Ito[1], T. Mitsuhashi[1], T. Nojima[1], and A. Tsukazaki[1,2]

[1]*Institute for Materials Research, Tohoku University, Sendai 980-8577, Japan.*

[2]*PRESTO, Japan Science and Technology Agency (JST), Chiyoda-ku, Tokyo 102-0075, Japan.*

#Correspondence to:   junichi.shiogai@imr.tohoku.ac.jp

Among the recently discovered iron-based superconductors[1-3], ultrathin films of FeSe grown on SrTiO$_3$ substrates have uniquely evolved into a high superconducting-transition-temperature ($T_C$) material[4-15]. The mechanisms for the high-$T_C$ superconductivity are ongoing debate mainly with the superconducting gap characterized with *in-situ* analysis for FeSe films grown by bottom-up molecular-beam epitaxy. Here, we demonstrate the alternative access to investigate the high-$T_C$ superconductivity in ultrathin FeSe with top-down electrochemical etching technique in three-terminal transistor configuration. In addition to the high-$T_C$ FeSe on SrTiO$_3$, the electrochemically etched ultrathin FeSe transistor on MgO also exhibits superconductivity around 40 K, implying that the application of electric-field effectively contributes to the high-$T_C$ superconductivity in ultrathin FeSe regardless of substrate material. Moreover, the observable critical thickness for the high-$T_C$ superconductivity is expanded up to 10-unit-cells under applying electric-field and the insulator-superconductor transition is electrostatically controlled. The present demonstration implies that the electric-field effect on both conduction and valence bands plays a crucial role for inducing high-$T_C$ superconductivity in FeSe.

Iron selenide (FeSe) is one of the iron-based superconductors ($T_C \sim 8$ K)[1-3], which has tetragonal PbO structure consisting of the layered atomic stacking of iron and selenium along *c*-axis. In the single or two unit-cell FeSe films grown by the state-of-the-art bottom-up molecular beam epitaxy (MBE)[4-15] on insulating $SrTiO_3$ (refs. 12-14) or Nb-doped $SrTiO_3$ (refs. 4-12 and 15), a large superconducting gap of about 20 meV has been proved by angle-resolved photoemission spectroscopy (ARPES)[4-10] and scanning tunneling microscopy (STM)[11]. Compared to the expected $T_C$ as high as 65 K (ref. 6) or 80 K (ref. 11) from the large gap energy evaluated by ARPES or STM, temperature dependence of resistivity gauges the consensus of the onset $T_C$ value about 40 K (refs. 11-14). However, a very recent report of superconductivity above 100 K detected by *in-situ* electrical resistance measurements for FeSe films grown on Nb-doped $SrTiO_3$ would make $T_C$ become widespread[15,16]. The mechanisms for the enhancement of superconducting gap energy in ultrathin FeSe are now discussed about strong electron-phonon coupling[10], variation of the electronic band structure at M point and Γ point[4], and charge transfer effect from substrates[5,13]. *Ex-situ* characterizations based on temperature dependence of resistivity regarding electric-field effect, magnetic-field effect, thickness dependence of $T_C$ and substrate material dependence are matters of interest for deep understanding the high-$T_C$ superconductivity in ultrathin FeSe studies.

In contrast to the *in-situ* bottom-up MBE technique, a classical electrochemical etching[17] is one of the best ways for top-down approach to the ultrathin FeSe films, because the uniform and fine tuning in thickness is possible with continuously monitoring electrical current originating from the electrochemical reaction. In our experiments, we employed this electrochemical etching approach in three-terminal transistor configuration, so-called an electric double layer (EDL) transistor (Fig. 1a)[18,19]. The EDL transistors were fabricated from 13- and 18-nm-thick FeSe films grown on insulating $SrTiO_3$ and MgO substrates, respectively, by pulsed-laser deposition (PLD) as described in Methods. The films and Pt gate electrode are covered with an ionic liquid (IL) of DEME-TFSI as schematically shown in Fig. 1a. The original concept of the EDL transistor involves an electrostatic charge carrier doping based on giant capacitance originating from a very narrow (< 1 nm) EDL width. Following this concept, various kinds of physical properties such as superconductivity[19], ferromagnetism[20,21] and insulator-metal transition in correlated electron systems[22,23] have been induced or controlled electrostatically. Another feature of EDL transistors is electrochemical reaction[24-26]. These two features of EDL transistor appear separately, depending on the temperature of IL and the applied gate voltage[24,25], with some boundary as schematically shown by the black broken line in Fig. 1b. In this study, we have applied the electrostatic doping effect with $V_G$ variation below the characteristic temperature (blue arrow in Fig. 1b) and the electrochemical etching to peel FeSe layers with tuning the sample temperature under $V_G$ applied constant (red arrow in Fig. 1b). After each etching sequence, systematic thickness dependence and field-effect of the superconductivity in FeSe films can be investigated in a single EDL transistor.

Temperature dependence of the channel sheet resistance ($R_s$-$T$ curve) and its superconducting properties were measured using a standard d. c. four-terminal method. In the initial thick films on SrTiO$_3$ and on MgO, the onset superconducting transition temperature $T_c^{on}$ = 8.2 K and 4.8 K are clearly observed by lowering temperature $T$ as shown in Fig. 1c (solid lines), which is consistent with the bulk properties of FeSe (ref. 3). No significant change in $R_s$-$T$ curves (broken lines) is observed with electrostatic doping via $V_G$ = 5 V at 220 K for the initial thick condition, implying that high-$T_C$ superconductivity is not induced at IL/FeSe interface (see Supplementary Information) at this temperature.

The high-$T_C$ superconductivity appears in both FeSe films grown on SrTiO$_3$ (sample A) and MgO (sample N) after a series of etching as shown in Fig. 2a and 2b, respectively. With carrying forward the etching sequence with applying $V_G$ = 5 V at about 245 K and $R_s$-$T$ measurements regularly under the constant $V_G$ (black arrow in Fig.

1b), the onset $T_C$ in normalized sheet resistance $R_s/R_s^{100K}$ is drastically enhanced from the bulk value to the higher one of $T_c^{on}$ = 43.3 and 39.0 K for sample A and N, respectively. The estimated $T_c^{on}$ from $R_s$-$T$ curves shown in Fig. 2a and 2b are summarized in Fig. 2c and 2d, respectively as a function of the averaged thickness, which is carefully estimated from the initial thickness and the etching rate in electrochemical reaction via the gate leakage current (see SI). Four implications are contained in these figures. The first is that the two-step transitions are observed in the intermediate etching region. Accompanying with the enhancement of metallic behavior, $R_s$ drops first at around $T$ = 30 K and second at 8 K and 4 K in the case of FeSe on SrTiO$_3$ and MgO, respectively, as a first signature of the appearance of high-$T_C$ superconducting region in the device. The systematic disappearance of the bulky low-$T_C$ region with decreasing thickness evidences our fine tuning of the thicknesses. Indeed, the zero resistance temperature of about 30 K implies that the FeSe layers are uniformly etched down to the one/two-unit-cells in our primitive devices. The second is that the high-$T_C$ superconductivity emerges at the ultrathin FeSe film on MgO thanks to the electric-filed effect in EDL transistor. Such high-$T_C$ superconductivity in FeSe on MgO has not been reported in *in-situ* MBE studies, which is probably due to no charge transfer from MgO to FeSe. The third is that the values of $T_c^{on}$ for both samples on SrTiO$_3$ and MgO are close to the highest values that have been reported in capped single-unit-cell FeSe films grown by MBE[12-14]. If the disorder at IL/FeSe interface limits $T_c^{on}$ to about 40 K as low as the capped FeSe, we need further improvement of interface quality to enhance $T_c^{on}$ toward 100 K (refs. 15 and 16). The forth is that the high-$T_C$ superconductivity emerges under a relatively thicker condition about 10-unit-cells, followed by the small thickness dependence. At present, it is not clear whether whole thick film region contributes high-$T_C$ superconductivity or not. We will discuss it later.

To further investigate the superconducting properties of the ultrathin FeSe film, we performed $R_s$-$T$ measurements under perpendicular magnetic fields $B$ for sample N on MgO. As shown in Fig. 2e, the tail of resistive transition is broadened by applying $B$ indicating the two dimensional nature of the FeSe film. Figure 2f shows the temperature dependence of perpendicular upper critical field $B_{c2}(T)$ derived from the midpoint temperature of the resistance drop $T_c(B)$ in $R_s$-$T$ curves in Fig. 2e. From the linear relation, which is consistent with the Ginzburg-Landau (GL) theory given as $B_{c2}(T) = \frac{\phi_0}{2\pi\xi^2(0)}(1 - T/T_c(0))$ with $\phi_0$ and $\xi(0)$ being the flux quantum and the GL coherent length at zero temperature, respectively, we obtained $\xi(0) = 2.0$ nm, which is around two times shorter than the bulk value[3].

Getting back to the thickness dependence of $T_c^{on}$ in Figs. 2c and 2d, we note that the first single step superconducting transition appears at the thickness of 9.6 nm and 5.9 nm for sample A and N, respectively. The appearance of high-$T_C$ superconductivity in the rather thicker conditions offers us to consider that how the electric-field effect contributes to the high-$T_C$ superconductivity. Here, we examined the effect of electrostatic doping on the superconductivity in both ultrathin and thicker conditions by different experimental schemes in the same device set-up as sample A. At first, samples B, C on SrTiO$_3$ and M on MgO were etched at $T = 245$ K to induce the superconductivity as shown in Fig. 3a, the top panel of Fig. 3c and Fig. 3d (red line data), respectively. The thicknesses of samples B, C and M are tuned to the one/two-unit-cell (the red square in Fig. 2c), 9.4 nm (the blue triangle in Fig. 2c) and 3.7 nm (the green circle in Fig. 2d), respectively, by monitoring the leakage current under the assumption that the same etching rate with sample A holds. After the detection of the high-$T_C$ superconducting behavior, we examined the electrostatic effect with removing and then applying $V_G$ at $T = 220$ K (along the blue arrow in Fig. 1b; Note that no electrochemical etching occurs at this temperature as mentioned above).

For sample B expectedly in the one/two-unit-cell condition, the initial insulating behavior was not recovered by removing $V_G$ from 5 V to 3 and 0 V as shown in Fig. 3b.

This result is a good evidence for the thickness reached close to one/two-unit-cell, because the high-$T_C$ superconductivity under $V_G = 0$ V is consistent with the result of previous reports on FeSe/SrTiO$_3$ with charge transfer solely from the SrTiO$_3$ substrate[8]. Interestingly, the increase of $T_c^{on}$ from 43.5 K to 46.3 K as a result of reduced $V_G$ implicates the over-doping effect by electrostatic doping. In contrast, in sample C under the thicker 9.4 nm condition, the superconductivity vanishes and the $R_s$-$T$ curve is back to the insulating behavior when $V_G$ is removed (blue line in middle panel of Fig. 3c). The reappearance of insulating behavior indicates the uniform etching without local ultrathin region providing a high-$T_C$ superconducting current pass. Then, application of $V_G = 5$ V recovers the superconductivity again (blue line in bottom panel of Fig. 3c), demonstrating a reversible control of insulator-superconductor transition by electrostatic mean. These series of results in sample C suggests that the observable thickness condition for high-$T_c^{on}$ superconductivity is widely expanded from one/two-unit-cell to above 10-unit-cells by the application of electric field. Additionally, $T_c^{on}$ in sample M decreases with reducing the accumulated charge as shown in Fig. 3d. Therefore, we suppose that the high-$T_C$ superconductivity in ultrathin FeSe on MgO is fully controllable by the application of electric-field effect.

Finally, we discuss possible mechanisms for the high-$T_C$ superconductivity in FeSe EDL transistor from the view point of the thickness dependence. The comparable values of $T_c^{on}$ as shown in Fig. 2c and 2d for SrTiO$_3$ and MgO suggests that the electric-field effect induces superconductivity at 40 K despite that one may expect the different electron-phonon coupling between FeSe film and substrate[10]. Concomitantly, the small thickness dependence of $T_c^{on}$ below 10-unit-cell indicates that an ultrathin condition, particularly single-unit-cell itself, does not probably play a crucial role to the increase of $T_C$ in ultrathin FeSe superconductivity. Nevertheless, the origin for the difference of critical thickness for SrTiO$_3$ (9.6 nm) and MgO (5.9 nm) is still not clear at present although it may link to the degree of charge transfer from substrates to FeSe. Naively, it is likely that the positive $V_G$ depletes holes in the valence band at the Γ point and accumulates electrons in the conduction band at the M point through the IL/FeSe interface, while the interface between IL and thick FeSe layer does not work as the high-$T_C$ superconducting channel (see Fig. 1c). As a consequence, the similar modification of the electronic band structure at the Γ and M point, as is proved by ARPES for the single-unit-cell FeSe on SrTiO$_3$ exhibiting large superconducting gap energy[5], may be realized by applying electric-field below the critical thicknesses on SrTiO$_3$ and MgO. However, we need further investigation about the modification of electronic band structures under $V_G$ application to verify our speculation.

By employing EDL transistor configuration, the high-$T_C$ superconductivity in ultrathin FeSe films has been investigated from the several aspects such as thickness and substrate material dependence. The application of electric-field effect inducing modification of electronic band structure would make a major contribution for $T_c^{on}$ of 40 K below 10-unit-cells regardless of substrate material. Our demonstration of the etching well controlled by the temperature of IL under applying $V_G$ shed light to the new aspect of the EDL transistor that is a useful approach to ultrathin films studies, expanding an opportunities to exemplify *ex-situ* the superconductivity of ultrathin FeSe grown on various insulating substrates[12,13,15,27].

**Methods**

**Sample fabrication.** The FeSe films were grown at 300 °C on insulating SrTiO$_3$ (001) and MgO (001) substrates by pulsed-laser deposition with as-delivered FeSe target (Kojundo Chemical Laboratory Co., Ltd), which was followed by the *in-situ* post annealing at 450 °C for 30 min. The growth rate with laser repetition rate of 5 Hz is approximately 17 nm/h, whose value is estimated by x-ray Laue fringes (See SI). After cutting the film into a 2 × 5 mm² rectangular piece, the channel and the four-terminal electrode (indium pads) were prepared to confine the current flow within the channel by scratching the film. Then, the electrodes were covered by silicone sealant to prevent them from chemical reaction with the ionic liquid. A platinum plate for the side gate electrode was placed next to the FeSe channel on the same chip and indium pads are attached on the gate electrode in the same way (See SI). Then, the FeSe channel and the gate electrode are covered by the ionic liquid, *N,N*-dethyl-*N*-methyl-*N*-(2-methoxyethyl)ammonium bis(trifluoromethanesulfonyl)imide (DEME-TFSI) to form electric double layer transistor configuration as shown in Fig. 1a. To suppress the degradation of the sample in the air, we carried out the EDL fabrication process as soon as possible. However, it usually took 6 hours before starting the measurement and the sample surface was inevitably exposed to the air about 2-3 hours.

**Electrical measurements.** The sheet resistance $R_s$ is measured by standard four-terminal method where a constant current of typically 0.5 µA is applied and the longitudinal voltage drop is measured by a nano-voltmeter. Temperature is controlled by dipping the sample into the liquid helium vessel in electrical measurements for samples A - C.

Sample temperature is monitored by a calibrated resistance thermometer (Cernox1050 resistor) inserted in the cupper block where the sample is mounted. The $R_s$-$T$ curve and magnetotransport properties for samples M and N are measured by Quantum Design Physical Property Measurement System (PPMS). To control the doping level electrostatically through EDL tuning, we fix the temperature at $T$ = 220 K during $V_G$ variation. In contrast, the occurrence of the electrochemical reaction has been checked by tuning $T$ within the range from 240 K to 250 K with keeping the $V_G$ = 5.0 V constant.

**Acknowledgments**

The authors thank S. Tanaka and T. Ouchi for experimental help and M. Kawasaki, Y. Iwasa and K. Fujiwara for stimulating discussions. This work is partly supported by Grant-in-Aid for Scientific Research on Innovative Areas (No. KAKENHI 22103004) from MEXT of Japan and Grant-in-Aid for Specially Promoted Research (No. KAKENHI 25000003) from the Japan Society for the Promotion of Science (JSPS).

**Author contributions**

J. S., T. N. and A. T. conceived and designed this research. Y. I. and J. S. performed thin films growth and electrical measurements. T. M. and T. N. prepared electrical measurement system for EDL transistor device configuration in liquid helium vessel. Y. I. and J. S. analyzed the data and all authors jointly discussed. J. S. and A. T. wrote the manuscript and all authors commented on the manuscript.

**Competing financial interests**

The authors declare no competing financial interests.

**Figure captions**

**Figure 1| Electrochemical etching and electrostatic doping in electric double layer transistor. a**, Schematic of the EDL transistor composed of FeSe films on insulating SrTiO$_3$ and MgO substrates. Side gate voltage $V_G$ is applied through the ionic liquid. The gate leakage current $I_G$ is measured by picoammeter, which is a measure for the degree of etching. **b**, Schematic diagram displaying the role of EDL transistor as a function of $V_G$ and temperature $T$. The broken line indicates a kind of boundary between electrostatic doping and electrochemical etching. **c**, The $R_s$-$T$ curves of sample A on SrTiO$_3$ and sample M on MgO for initial thick condition when $V_G$ = 0 V (solid lines) and 5 V (broken lines) are applied.

**Figure 2| Thickness tuning of FeSe on SrTiO$_3$ and MgO via electrochemical etching. a,b,** Sheet resistance normalized to the value at $T$ = 100 K, $R_s/R_s^{100K}$, as a function of temperature $T$ for continuing etching processes *i.e.*, different film thickness (*d*), for sample A on SrTiO$_3$ and for sample N on MgO, respectively. The onset superconducting transition temperature $T_c^{on}$ is defined by the intersection of the broken two linear extrapolation lines. **c,d**, $T_c^{on}$ for samples A and N shown in Fig. 2a and 2b, respectively, are collected as a function of averaged thickness for the electric-field

induced high-$T_C$ (E-induced; filled black circles) and the initial bulk-like low-$T_C$ (open black circles) conditions. A vertical broken line corresponds to the thickness for the first data showing single step transition in $R_s$-$T$ curves. The data for samples B, C and M after etching and at $V_G$ = 5 V are indicated by red square, blue triangle and green circle, respectively. **e**, Perpendicular magnetic field dependence of $R_s$-$T$ curves of sample N after etching down to $d$ = 0.5 nm measured at $B$ = 0, 1, 3, 5, 7 and 9 T. **f**, Temperature dependence of the upper critical field $B_{c2}(T)$ derived from the midpoint of the resistance drop in Fig. 2e. Red line corresponds a linear fitting line based on the GL theory.

**Figure 3 | Electrostatic tuning of superconductivity. a**, Normalized sheet resistance $R_s/R_s^{100K}$ as a function of temperature $T$ for sample B in the initial 13-nm-thick state with applied $V_G$ = 5 V (gray solid line) and after etching down to one unit-cell with $V_G$ = 5 V (red solid line). **b**, The $R_s$-$T$ data for sample B after thinning to the 0.5 nm. $V_G$ is reduced from $V_G$ = 5 V (red), 3 V (green), to 0 V (gray). **c**, The reversible variation in $T$ dependence of $R_s/R_s^{100K}$ for sample C against $V_G$. (Top panel) The initial insulating (13 nm thick) states with $V_G$ = 5 V (gray solid line) and the superconducting behavior after the electrochemical etching to 9.4-nm-thick condition at $V_G$ = 5V (blue). (Middle panel) Blue solid line shows $R_s/R_s^{100K}$ after removing $V_G$ to 0 V. (Bottom panel) Blue solid line

shows $R_s/R_s^{100K}$ with the second application of $V_G = 5$ V. **d**, The $R_s$-$T$ curves measured for sample M under 3.7 nm condition. Transition behavior is presented as reducing charges from red solid line (charged at $V_G = 5$ V) to blue solid line (discharged at $V_G = 0$). Triangles correspond to the $T_c^{on}$.

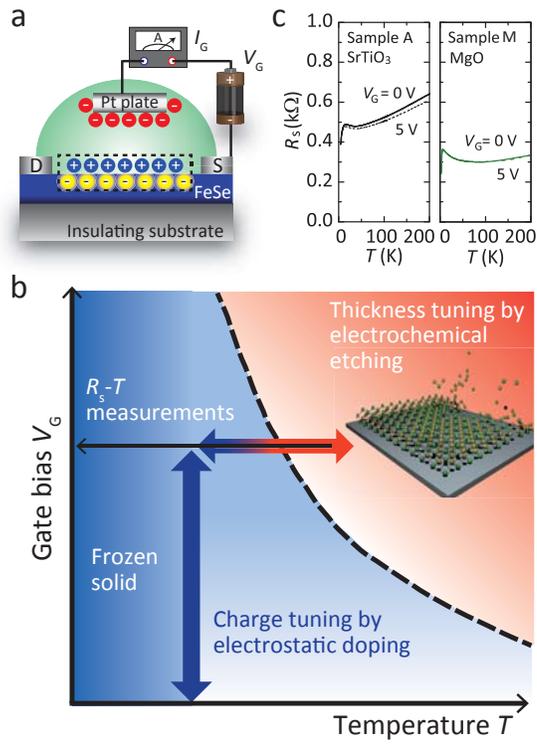

Figure 1 J. Shiogai *et al.*,

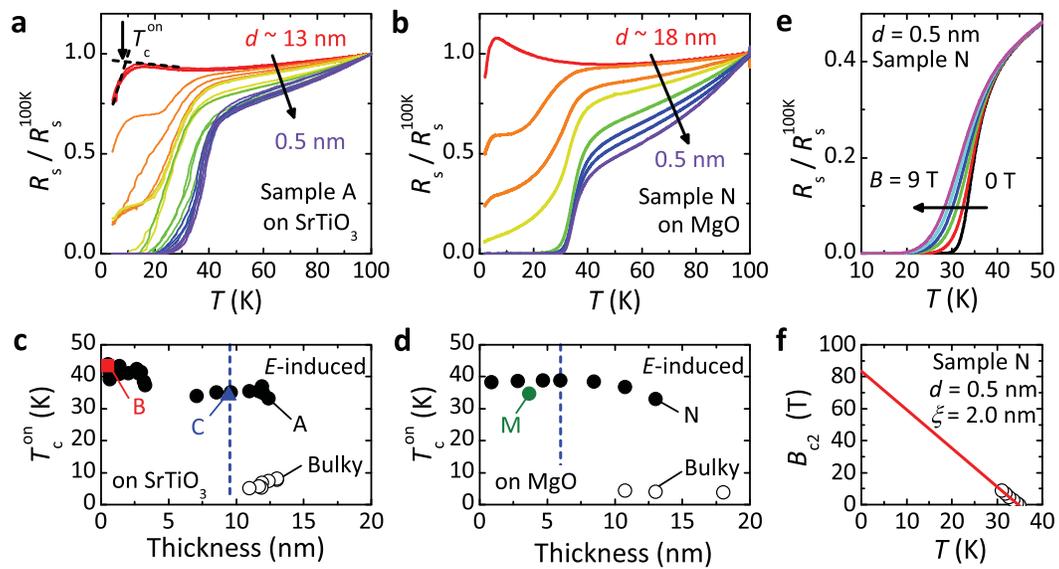

Figure 2 J. Shiogai *et al.*,

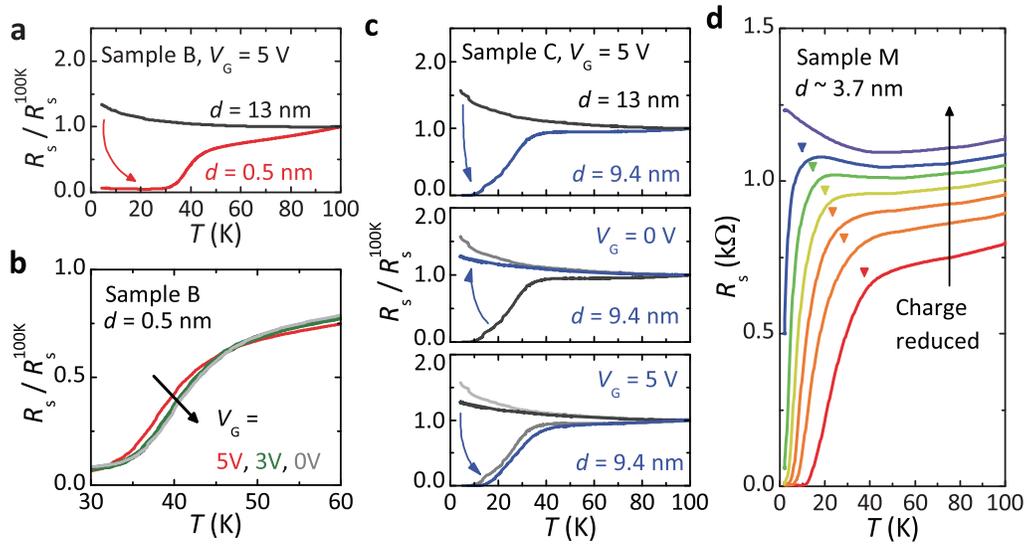

Figure 3 J. Shiogai *et al.*,